\pdfoutput=1
\documentclass{sig-alternate}
\usepackage{cite}
\usepackage{url}

\PassOptionsToPackage{hyphens}{url}\usepackage{hyperref}

\begin{document}



\conferenceinfo{MM '16}{October 15--19, 2016, Amsterdam, The Netherlands}

\newcommand{\zhi}[1]{\textcolor{blue}{{\bf} #1}}


\title{Drone Streaming for Virtual Tour:\\A Wi-Fi Grid Aggregation Approach}

\numberofauthors{1}
\author{
\alignauthor
Chenglei Wu, Zhi Wang, Shiqiang Yang\\
       \affaddr{Tsinghua University}\\
}
\maketitle

\begin{abstract}

    To provide a live, active and high-quality virtual touring streaming experience, we propose an unmanned drone stereoscopic streaming paradigm using a control and streaming infrastructure of a 2.4GHz Wi-Fi grid. Our system allows users to actively control the streaming captured by a drone, receive and watch the streaming using a head mount display (HMD); a Wi-Fi grid is deployed across the remote scene with multi-channel support to enable high-bitrate streaming broadcast from the drones. The system adopt a joint view adaptation and drone control scheme to enable fast viewer movement including both head rotation and touring. We implement the prototype on Dji M100 quadcopter and HTC Vive in a demo scene.


\end{abstract}


\section{Introduction} \label{sec:introduction}

The rapid development of video coding, streaming and wireless technologies has enabled a new type of touring: virtual tour, in which a ``tourist'' is able to receive live video streaming from a remote location, to achieve a virtual touring experience. In previous studies, such experience is usually provided in a \emph{staled} and \emph{passive} manner: the user only receives either a staled video stream of the remote scene (e.g., Google Street View) captured previously, or a live stream captured by others (e.g., Youtube $360^{\circ}$ Videos) that cannot be really controlled by the viewer.

\begin{figure}[!th]
    \centering
    \includegraphics[width=.9\linewidth]{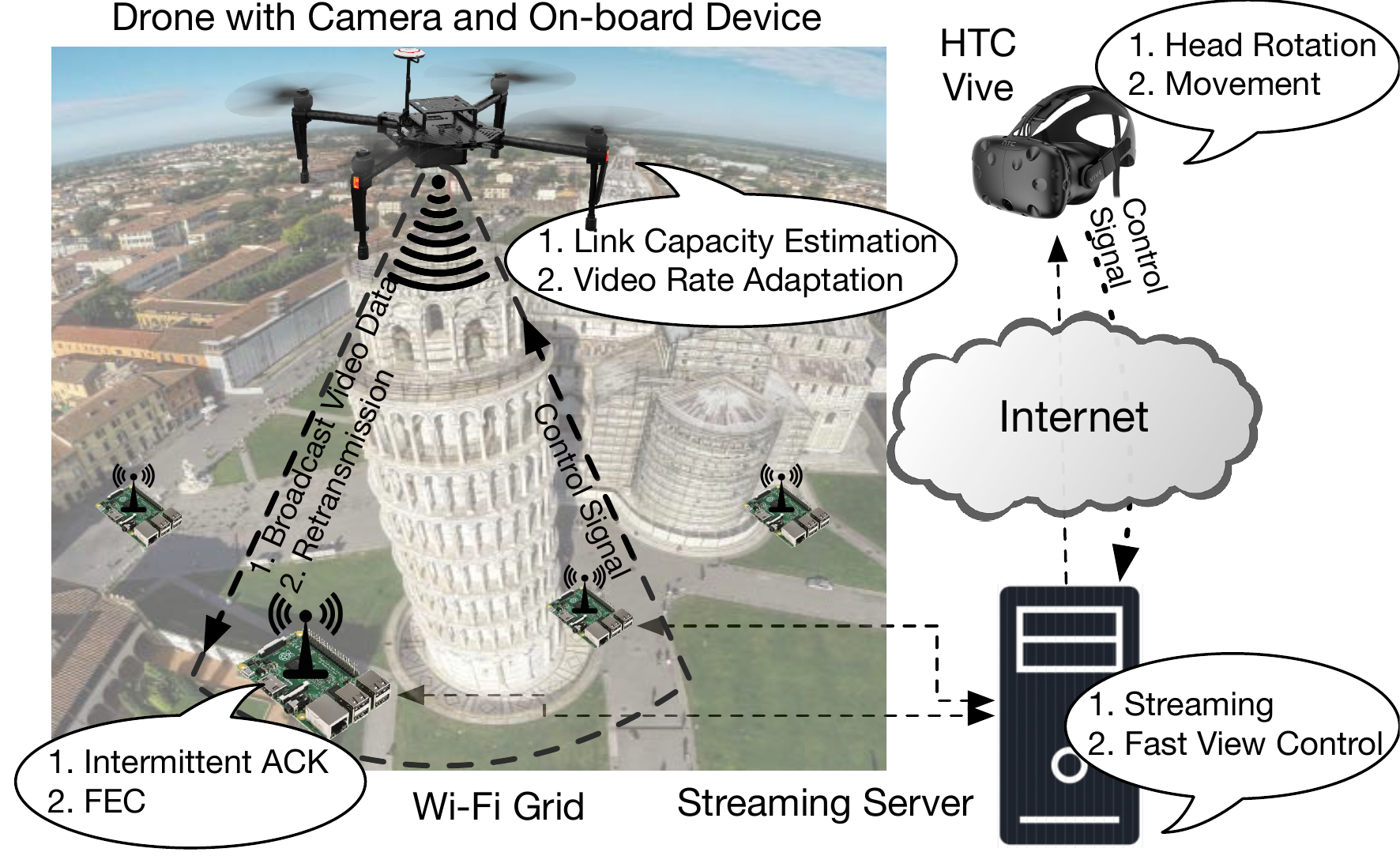}
    \caption{System Overview}
    \label{fig:overview}
\end{figure}

It is thus still appealing to enable live, active and high-quality virtual touring streaming experience. The release of virtual reality (VR) headset (e.g., HTC Vive) and programmable unmanned drone (e.g., Dji M100) brings more possibilities to the field of virtual tour: Watching stereoscopic video taken by a drone whose movement is controlled by a viewer waring a VR headset, as illustrated in Fig.~\ref{fig:overview}.

Enabling drone streaming for virtual tour is however challenging.

First, long-distance touring and high-quality streaming is challenging. Former studies that use drones to achieve virtual touring usually use a Wi-Fi transmitter and a receiver \cite{wifibroadcast, mirk2015virtual} for streaming from the drone to the receiver, and both the flight range and transmission bandwidth are limited. For example, the transmission range of a state-of-the-art product, Dji Lightbridge 2, is under $1.7$km with a video stream up to $1080$p@$30$fps\footnote{ http://www.lemondrone.com/blog/dji-lightbridge-explained-what-is-lightbridge-2/ \label{footnote:lightbridge}}. The situation is further exacerbated under obstruction and interference, e.g., the transmission quality degrades significantly in the urban areas.


Second, active streaming control based on a viewer's head rotation and movement is challenging. Improper drone control not only affects the touring experience but also cause severe problems (e.g., fatigue or nausea). To provide immersive virtual reality experience, users should be able to control the drone directly with their head rotation and movement, and the drone has to respond quickly to the viewer's behaviors.

To address these issues, we propose an unmanned drone stereoscopic streaming paradigm using a control and streaming infrastructure of a $2.4$GHz Wi-Fi grid. Our system features: (1) Wi-Fi grid infrastructure: a Wi-Fi grid is deployed across the remote scene with multi-channel support to enable high-bitrate streaming broadcast from the drones; (2) Joint active control and immersive viewing UI: we let users actively control the streaming captured by a drone, receive and watch the streaming taken using a head mount display (HMD); (3) Fast view control: a joint view adaptation and drone control scheme is proposed to enable fast viewer movement including both head rotation and movement.

\section{System Overview} \label{sec:overview}

Three modules are designed and implemented in our prototype as shown in Fig.~\ref{fig:overview}. The first one is an on-board device which captures stereoscopic video and \emph{broadcasts} it using Wi-Fi channels. The second one is a Wi-Fi grid infrastructure which is composed of densely deployed Wi-Fi receivers to receive the broadcast streaming from the drone. The last one is a streaming server which streams reorganized video data uploaded by Wi-Fi grid to user's HMD and enables drone control with user's body movement.

\subsection{On-board Device}
The on-board device is an embedded linux device with multiple Wi-Fi antennae\footnote{
It communicates with the drone's flight controller
over a direct serial connection to monitor and control aircraft flight behavior.}.
It takes stereoscopic video using the camera module carried by the drone and
broadcasts video data to Wi-Fi channels using UDP to mitigate
the delay caused by Wi-Fi association and acknowledgement,
i.e., UDP packets are broadcast via raw packets\cite{multilayermulitcast}.
It has a video rate controller to control the video capturing bitrate
according to the estimated link capacity,
which is determined by the UDP throughput.

\subsection{Wi-Fi Grid Infrastructure}
The Wi-Fi grid is a densely deployed grid of Wi-Fi receivers. At least
two or three of the receivers are in the transmission range of the on-board device.
A receiver is also an embedded linux device with multiple Wi-Fi antennae.
It listens on the selected Wi-Fi channels to capture the video data packets
broadcasted by the on-board device and performs forward error correction
(FEC)\cite{nafaa2008forward} on the received data to repair lost or corrupted packets.
It drops overdue packets and uploads useful data to the streaming server.
If more than 3 receivers are in range of the drone, only the closet 3 receivers
based on GPS coordinates will upload video data.
The closet receiver sends ACK at a frequency of once every ten
received packets to reduce the acknowledgment overhead
and feedbacks to the on-board device for bandwidth estimation.

\subsection{Streaming Server}
\subsubsection{Joint Active Control \& Immersive Viewing UI}
The streaming server receives video data uploaded by several Wi-Fi receivers
and re-organizes them to stream to viewer's HMD.
User's head rotation is tracked by headset to control camera's capture angle
by rotating the camera vertically and direction by rotating the drone horizontally.
User's movement is tracked by motion tracking sensors to control drone's
position. These movements are transformed into control signal and delivered via the
streaming server to Wi-Fi grid. The receiver that is closet to the drone
will emit the signal over Wi-Fi channel to the
on-board device that controls the drone directly.

\subsubsection{Fast View Control}
To compensate the transmission delay between drone and HMD, we proposed a joint
view adaptation and drone control scheme, in which we present only a part of
received image to user (display window).
When the user rotates her/his head, we move the display window to the viewing direction first,
followed by the drone and camera's movement.

\section{Demonstration} \label{sec:demo}
\begin{figure}
    \centering
    \includegraphics[width=\linewidth]{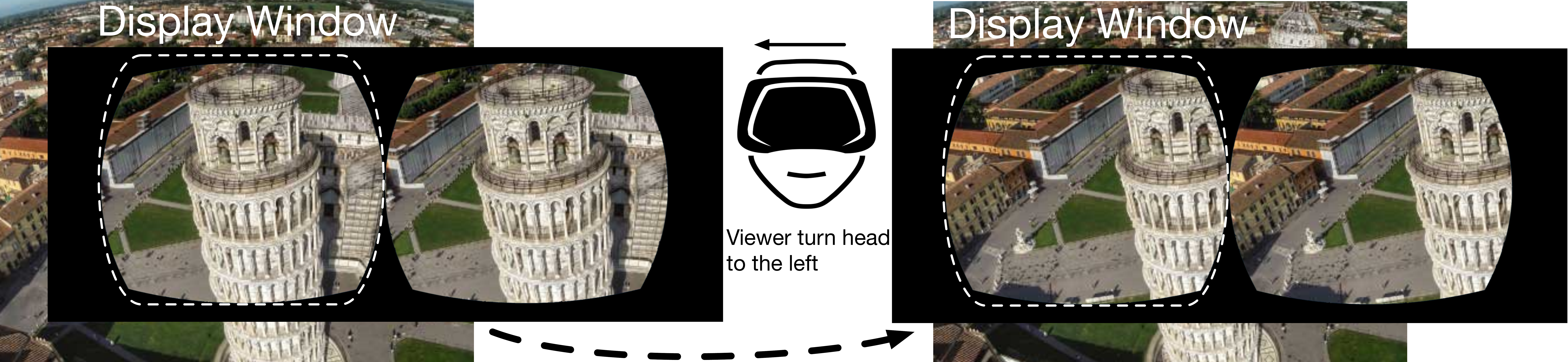}
    \caption{Example of Rendered View}
    \label{fig:rendered_view}
\end{figure}

We implemented the on-board device and Wi-Fi grid receiver using Raspberry
Pi 2. Each device had two Wi-Fi antennae to aggregate bandwidth. The Wi-Fi antennae
on the aircraft included a power amplifier to achieve a better range.
The drone we chose was Dji M100 programmable quadcopter which provides SDK to control
the behavior of drone and gimbal over a direct serial connection.
We deployed the receiver device every 500 meter to cover the whole area and connected
them to a PC over a local area network.
The HMD device we chose was HTC Vive virtual reality headset which
has a pair of laser position sensors and can track the user's movement in a 4.6 meters by 4.6 meters space.
User was able to watch the video captured by the drone with HMD and controlled the
drone and camera by rotating his head or moving his position (Fig. \ref{fig:rendered_view}).

\section{Discussion and Summary} \label{sec:conclusion}

We proposed an unmanned drone stereoscopic streaming paradigm using an on-board device, a Wi-Fi grid as control and streaming infrastructure, and a VR headset to provide virtual touring experience. We implemented the system to support high-bitrate streaming with bandwidth aggregation and fast active drone and view control. User are able to watch the stereoscopic video through HMD and control the drone with body movement.
In future, we plan to conduct intensive researches to enable more complicated touring experience using the Wi-Fi grid infrastructure.

\bibliographystyle{abbrv}
\bibliography{main}
\end{document}